\documentclass{hep99}
\begin{document}
\begin{titlepage}
\setcounter{page}{0}
\topmargin 2mm
\begin{flushright}
{\tt hep-ph/9910397}\\
{RM3-TH/99-12}\\
\end{flushright}
\begin{center}
{\bf TESTS OF THE STANDARD MODEL AT HERA} 
\vskip 12pt
{Stefano Forte\footnote[1]{On leave from INFN, Sezione di Torino, Italy}}\\
\vskip 6pt
{\em INFN, Sezione di Roma III,\\ 
via della Vasca Navale 84, I-00146 Rome, Italy}\\
\vskip 120pt
\end{center}
\centerline{\bf ABSTRACT}
\bigskip\noindent
{\advance\leftskip by 36truept\advance\rightskip by 36truept
We review how experimental data collected at the HERA lepton-hadron
collider have improved our theoretical and phenomenological
understanding of the standard model, and specifically of its QCD sector.
\smallskip
}

\begin{center}
\vskip 130pt
{Invited talk given at the\\\smallskip {\em EPS Conference on High
Energy Physics}\\ Tampere, Finland,
July 1999}\\
\smallskip
{\em to be published in the proceedings}\\

\end{center}
\bigskip
\vfill
\begin{flushleft}
{October 1999}
\end{flushleft}
\end{titlepage}

\def\Im{\,\hbox{Im}\,}
\def\Re{\,\hbox{Re}\,}
\def\tr{\,{\hbox{tr}}\,}
\def\Tr{\,{\hbox{Tr}}\,}
\def\det{\,{\hbox{det}}\,}
\def\Det{\,{\hbox{Det}}\,}
\def\neath#1#2{\mathrel{\mathop{#1}\limits_{#2}}}
\def\ker{\,{\hbox{ker}}\,}
\def\dim{\,{\hbox{dim}}\,}
\def\ind{\,{\hbox{ind}}\,}
\def\sgn{\,{\hbox{sgn}}\,}
\def\mod{\,{\hbox{mod}}\,}
\def\apm#1{\hbox{$\pm #1$}}
\def\epm#1#2{\hbox{${\lower1pt\hbox{$\scriptstyle +#1$}}
\atop {\raise1pt\hbox{$\scriptstyle -#2$}}$}}
\def\neath#1#2{\mathrel{\mathop{#1}\limits_{#2}}}
\def\gsim{\mathrel{\rlap{\lower4pt\hbox{\hskip1pt$\sim$}}
    \raise1pt\hbox{$>$}}}         
\def\eg{{\it e.g.}}
\def\ie{{\it i.e.}}
\def\viz{{\it viz.}}
\def\etal{{\it et al.}}
\def\rhs{right hand side}
\def\lhs{left hand side}
\def\toinf#1{\mathrel{\mathop{\sim}\limits_{\scriptscriptstyle
{#1\rightarrow\infty }}}}
\def\tozero#1{\mathrel{\mathop{\sim}\limits_{\scriptscriptstyle
{#1\rightarrow0 }}}}
\def\frac#1#2{{{#1}\over {#2}}}
\def\half{\hbox{${1\over 2}$}}\def\third{\hbox{${1\over 3}$}}
\def\quarter{\hbox{${1\over 4}$}}
\def\smallfrac#1#2{\hbox{${{#1}\over {#2}}$}}
\def\pbp{\bar{\psi }\psi }
\def\vevpbp{\langle 0|\pbp |0\rangle }
\def\as{\alpha_s}
\def\tr{{\rm tr}}\def\Tr{{\rm Tr}}
\def\eV{{\rm eV}}\def\keV{{\rm keV}}
\def\MeV{{\rm MeV}}\def\GeV{{\rm GeV}}\def\TeV{{\rm TeV}}
\def\MS{\hbox{$\overline{\rm MS}$}}
\def\blackbox{\vrule height7pt width5pt depth2pt}
\def\matele#1#2#3{\langle {#1} \vert {#2} \vert {#3} \rangle }
\def\VertL{\Vert_{\Lambda}}\def\VertR{\Vert_{\Lambda_R}}
\def\Real{\Re e}\def\Imag{\Im m}
\def\bp{\bar{p}}\def\bq{\bar{q}}\def\br{\bar{r}}
\catcode`@=11 
\def\slash#1{\mathord{\mathpalette\c@ncel#1}}
 \def\c@ncel#1#2{\ooalign{$\hfil#1\mkern1mu/\hfil$\crcr$#1#2$}}
\def\lsim{\mathrel{\mathpalette\@versim<}}
\def\gsim{\mathrel{\mathpalette\@versim>}}
 \def\@versim#1#2{\lower0.2ex\vbox{\baselineskip\z@skip\lineskip\z@skip
       \lineskiplimit\z@\ialign{$\m@th#1\hfil##$\crcr#2\crcr\sim\crcr}}}
\catcode`@=12 
\def\twiddles#1{\mathrel{\mathop{\sim}\limits_
                        {\scriptscriptstyle {#1\rightarrow \infty }}}}

\def\NCA{\em Nuovo Cimento}
\def\NIM{\em Nucl. Instrum. Methods}
\def\NIMA{{\em Nucl. Instrum. Methods} A}
\def\NPB{{\em Nucl. Phys.} B}
\def\PLB{{\em Phys. Lett.}  B}
\def\PRL{\em Phys. Rev. Lett.}
\def\PRD{{\em Phys. Rev.} D}
\def\ZPC{{\em Z. Phys.} C}

\def\st{\scriptstyle}
\def\sst{\scriptscriptstyle}
\def\mco{\multicolumn}
\def\epp{\epsilon^{\prime}}
\def\vep{\varepsilon}
\def\ra{\rightarrow}
\def\ppg{\pi^+\pi^-\gamma}
\def\vp{{\bf p}}
\def\ko{K^0}
\def\kb{\bar{K^0}}
\def\al{\alpha}
\def\ab{\bar{\alpha}}
\def\be{\begin{equation}}
\def\ee{\end{equation}}
\def\bea{\begin{eqnarray}}
\def\eea{\end{eqnarray}}
\def\CPbar{\hbox{{\rm CP}\hskip-1.80em{/}}}
\setlength\textwidth{39pc}
\setlength\textheight{54pc}

\title{Tests of the Standard Model at HERA}

\author{Stefano Forte$^\dag$}
%

\address{INFN, Sezione di Roma III,\\via della Vasca Navale 84,
I-00146 Roma, Italy}

\abstract{We review how experimental data collected at the HERA lepton-hadron
collider have improved our theoretical and phenomenological
understanding of the standard model, and specifically of its QCD sector. 
} 

\maketitle

\fntext{\dag}{On leave from INFN, sezione di Torino, Italy}

\section{Factorization}
HERA is unique as a lepton-hadron collider. The
measurement of lepton-hadron scattering cross sections allows for
detailed tests of the standard model thanks to the factorization
property of many hard QCD processes. The perturbative 
computation of the hard elementary process, and in particular strong
radiative corrections to it, allows a detailed test of both the
electroweak and strong sectors of the theory. Whereas of course 
electroweak tests are generally not competitive with the cleaner
setting of lepton colliders such as LEP, QCD tests at HERA allow
reaching kinematical regions which could not be attained in
fixed--target experiments, 
while offering a cleaner setting in comparison to hadron
colliders such as the Tevatron.

{\it 1.1 The electroweak subprocess}\\
The recent determination of the charged-current contribution to the
cross-section over a wide enough range of $Q^2$ allows a simultaneous
extraction of the Fermi constant $G_F$  and the $W$ mass\cite{wmass}; 
the ZEUS collaboration for instance gets
$M_W=80.9\epm{4.9}{4.6}\,(\hbox{stat.})\epm{5.0}{4.3}\,(\hbox{syst.})
\epm{1.3}{1.2}\,(\hbox{pdf.}).$
Notice that the last source of error is due to the choice parton
distributions. Assuming the correctness of the standard model and the
ensuing relation between
$M_W$ and $G_F$ leads to a value $M_W=80.5\pm 0.4$,  much better  than
the above model independent one. Even though this is inevitably not
competitive with the LEP determination, it provides a nice consistency test.

{\it 1.2 Testing QCD {\it vs.} using QCD}\\
Because of the
overwhelming success of perturbative QCD\cite{qcdrev}, fundamental
tests of its
correctness 
are these days  relatively  less
interesting in comparison to  the precise
determination of its free parameters. 
In practice, this includes  the strong coupling $\alpha_s$ but also
all quantities which can
not be calculated perturbatively, such as parton distributions. 
An accurate knowledge of these quantities is a necessary input to the
determination of any hadronic process, and is thus a
crucial ingredient for e.g. LHC physics. 
An accurate determination of QCD backgrounds to
new physics is also needed. Recent progress involves
more accurate determinations of parton distributions
(see Sect.~2), and
widening the perturbative domain, by learning
how to treat processes with many large scales (Sect.~3) and extending
factorization theorems to less inclusive processes (Sect.~4).

\section{Structure functions and parton distributions}
A striking success of perturbative QCD is the
prediction of scaling violations of structure functions. The proton structure
function $F_2(x,Q^2)$ in particular is measured at HERA for values of $Q^2$
extending up to $10^4$~GeV$^2$ and $1\over x$ up to $10^5$. Detailed
analyses\cite{f2fit} show excellent agreement between the data
and the next-to-leading order QCD prediction throughout
this region. Note that even though comparing data with theory 
requires a fit of parton distributions, the scale
dependence is then entirely predicted. This is to be contrasted with
model parametrizations, where the full $x,Q^2$ dependence of the
data is fitted by a given functional form.
Two features of these analyses\cite{f2fit} are worth
noticing. First, excellent agreement is obtained
with a value of the strong coupling fixed at  the
world average $\alpha_s(M_z)=0.118$, in contrast to earlier
indications that $\alpha_s$ from scaling violations should be smaller.
Also, contrary to expectations, next-to-leading order 
scaling violations  agree very
well with the data even at the boundaries of the kinematic region, in
particular at moderate $Q^2$, very large, and very small $x$ (see Sect.3). 

{\it 2.1 Errors on parton distributions}\\
The succesful description of structure functions and their scaling
violations within QCD
suggests that they can be reliably used to  determine parton
distributions.
Indeed, these  data
provide the strongest constraints on
current parton sets\cite{pdfs},
while less inclusive data (Sect. 3.2) give
additional constraints. 

Currently available parton sets do not
come equipped with errors. However, a recent study\cite{mrstwz} 
shows that if the W and Z production cross-section at Tevatron is
computed using different parton sets, the variation of the results 
is already comparable to the experimental errors (see also the
determination of $M_W$ above). 
However, independent
parton determinations share many theoretical assumptions,
and simply varying the pdf cannot provide a reliable error estimate. 
A better estimate can be obtained
by scanning the parameters of a given set\cite{mrstwz,cteqerr}, but 
the outcome then cannot be folded into subsequent analyses. 
An interesting suggestion\cite{bmc} 
to overcome these problems is based on the idea of
giving the results as a probability functional ${\cal P}[f]$
(rather than a fixed
parameterization), which can then be determined by Bayesian
inference in a monte carlo approach. The result can then be ported to
subsequent calculations. Determinations of pdfs based on this approach are
currently under way\cite{qf}. 

{\it 2.2 PDFs from non-inclusive processes}\\
Specific processes can provide stronger constraints on individual pdfs
than global fits based on structure function data. Such information
will be copiously collected by future experiments such as COMPASS. 
A recent example is the determination of
the gluon distribution\cite{gdj} from the dijet cross-section through its 
photon-gluon fusion component.  
Another example is the determination of the flavor
asymmetry in the nucleon sea $\bar d(x)-\bar u(x)$~\cite{udher}, where 
parent current quarks are identified by their preferred fragmentation
by tagging mesons in the final state. Although this measurement is not
competitive with that from Drell-Yan\cite{uddy}, similar, more refined 
measurements in future experiments could provide valuable constraints.
The relevance of such measurement is highlighted by the fact that
insufficient knowledge of the gluon distribution and the flavor
asymmetry of the sea had been respectively suggested\cite{cteqerr,np} 
as possible
explanations of the excesses of high $p_T$ jets at the Tevatron and of
high $Q^2$ events at HERA which had been initially interpreted as
possible indications of new physics. 
\section{QCD at small $x$}
Because the total center-of-mass energy of $\gamma^*-p$ collision is
$W^2={1-x\over x} Q^2$, the small-$x$ region of deep-inelastic
scattering has been accessed for the first time at HERA.

{\it 3.1 Double Asymptotic Scaling}\\
Perturbative QCD at leading and next-to leading order
predicts\cite{dgptwz,das} 
that
the gluon distribution, and thus the structure function $F_2(x,Q^2)$ at
small $x$ and large $Q^2$ should asymptotically depend on the single
variable $\sigma=\sqrt{\ln x \ln \ln Q^2}$, exponentially rising with
$\sigma$ with slope $\gamma=\sqrt{2 C_A/\beta_0}$ (double scaling).

This prediction is
beautifully borne out by the HERA data, contrary to the pre-HERA
expectation that the rise should be quenched by non-perturbative
effects, or substantially modified by higher order corrections (see
Sect.~3.2). The predicted and observed slopes
agree to about $10\%$, which is the size of the expected
subasymptotic corrections in the HERA region\cite{test}. 
This tests the fundamental nonabelian
nature of the gauge interaction.
The largeness and universality of small $x$ scaling violations and
the success of the NLO computation at HERA suggest that this is an
optimal region to determine $\alpha_s$, since the dependence on the
parton distributions will be minimal. This is borne out by a
preliminary analysis based on 1995 HERA data\cite{assx}. An extraction
based on current data would be very competitive with other recent determinations.
 
{\it 3.2 Small $x$ scaling violations}\\
The  double scaling rise of $F_2$ is driven by the 
(rightmost) simple pole
of
the LO and NLO anomalous dimensions. 
However, higher order contributions are known to display higher
order poles, and might be expected to modify double scaling. 
The coefficients of the leading singularities to all orders in $\alpha_s$
in the gluon sector can be extracted from the BFKL equation\cite{ft},
and are also known in the quark sector\cite{qad}. If a summation
of such contributions is included in the computation of
scaling violations, the agreement with the HERA data deteriorates\cite{sxph}.
Likewise, the subleading correction to these
coefficients can be extracted\cite{flad} from
the recently determined\cite{fl} subleading corrections to the BFKL
kernel. They are extremely large (and negative), 
and their ratio to the leading coefficient grows linearly with the
perturbative order. This perhaps explains the failure of the LO
summation, but makes the success of the unimproved NLO computation
all the more puzzling.

It can be shown that this bad large-order behavior of the summation of
small $x$ corrections to anomalous dimensions is generic, and can be
removed by a suitable reorganization of small $x$ perturbation
theory\cite{sxres}.
The reorganized expansion has a stable asymptotic
behavior, but still leads to large subasymptotic corrections with
a poorly
behaved perturbative expansion. It turns out that these
problem can in large part be cured by imposing suitable matching
conditions between large and small $x$ expansions, in particular
embodying momentum conservation. The success of standard two-loop
evolution can thus be accommodated within this framework\cite{abf}. 

Several more ways of dealing with the bad behavior of the small $x$
expansion have been suggested, based on various resummations of
formally subleading corrections\cite{flres}. 
It remains to be seen whether any of these approaches can lead to
succesful phenomenology. A firmer grasp of the pertinent phenomenology
will be required in order to reliably evolve parton distributions to
the large $Q^2$, small $x$ region relevant for LHC phenomenology.

{\it 3.3 Energy evolution}\\
The summation of small $x$ contributions to scaling violations
corresponds to a summation of leading $\ln {1\over x}$ contributions
to the deep-inelastic cross section. Such a summation can be
more directly obtained by solving an evolution equation in $1\over x$,
 i.e. in in the CM energy of the process (BFKL
equation). This is relevant for
processes where there is considerable
energy evolution  but little $Q^2$ evolution, such as
deep-inelastic forward
jet production, where the jet transverse
energy and photon vituality are similar, $k^2_T\sim Q^2$, but the
momentum fraction carried by the jet is large 
while Bjorken $x\ll 1$\cite{munav}. One expects then
a  resummation of $\ln {1\over x}$ to afford better phenomenology
than the usual $\ln Q^2$ resummation. This expectation is not
borne out by recent HERA data\cite{fjets}, which appear to be
adequately described by standard $Q^2$ evolution, provided only
one allows for a resolved photon component whenever $k^2_T\gsim Q^2$.

\section{Diffraction and leading hadrons}
Leading hadron processes are defined by tagging a hadron in the target
fragmentation region. Diffractive processes are then leading
proton (LP) events with  the further requirement that the LP
carries a large fraction of the incoming
hadron's momentum, i.e.  with a
rapidity gap between the LP and the remnant of the
final state. Surprisingly, diffractive events at HERA make
up for as much as 10\% of the total structure function $F_2$. An
understanding of diffractive p-p events is important because they 
are an important background to standard Higgs production.

Within perturbative QCD the leading hadron
cross-section can be proven to 
satisfy a factorization theorem\cite{dfac} which allows expressing it
in terms of a hard coefficient function and a fracture
function\cite{frac}, defined as the differential component of the
standard structure function in the presence of a leading hadron with
the given kinematics. Fracture functions satisfy the standard QCD
evolution equations. A phenomenological analysis of HERA diffractive
and leading proton data\cite{lpphen} shows that the predicted
universality and scale dependence of
fracture functions are well reproduced. The problem of 
computing (rather than fitting) fracture functions 
from first principles has attracted considerable theoretical 
attention\cite{diff}.

\section{Conclusion}
HERA has played for QCD a similar role as LEP for the
electroweak sector of the standard model. In general, perturbative
computations lead to excellent phenomenology; this however seems to
happen even beyond the regions were  one might expect it.

\noindent
{\bf Acknowledgement:} I thank G.~Passarino for convening this lively
session and for inviting me to put HERA physics in the context of the
standard model, and G.~Altarelli and R.~Ball for useful comments.

\end{document}